\documentclass[numberedappendix]{emulateapj}

\begin{document}


\bibliographystyle{apj}

\shorttitle{Polarized Submm Disk Limits}

\shortauthors{Hughes et al.}

\slugcomment{Accepted for Publication in ApJ: September 6, 2009}
\title{
Stringent Limits on the Polarized Submillimeter Emission from Protoplanetary 
Disks
}

\author{A. Meredith Hughes\altaffilmark{1},
David J. Wilner\altaffilmark{1},
Jungyeon Cho\altaffilmark{2,3},
Daniel P. Marrone\altaffilmark{4,5},
Alexandre Lazarian\altaffilmark{3},
Sean M. Andrews\altaffilmark{1,6},
Ramprasad Rao\altaffilmark{7}
}
\altaffiltext{1}{Harvard-Smithsonian Center for Astrophysics, 60 Garden Street, Cambridge, MA 02138; mhughes, dwilner, sandrews$@$cfa.harvard.edu}
\altaffiltext{2}{Dept. of Astronomy and Space Science, Chungnam National Univ., Daejeon, Korea; jcho@cnu.ac.kr}
\altaffiltext{3}{University of Wisconsin, Department of Astronomy, 475 N. Charter St., Madison, WI 53706; lazarian@astro.wisc.edu}
\altaffiltext{4}{The University of Chicago, The Kavli Institute for Cosmological Physics, 933 East 56th Street, Chicago, IL 60637; dmarrone@oddjob.uchicago.edu}
\altaffiltext{5}{Jansky Fellow}
\altaffiltext{6}{Hubble Fellow}
\altaffiltext{7}{Institute of Astronomy and Astrophysics, Academia Sinica.  7F of Condensed Matter Sciences and Physics Department Building, National Taiwan University.  No.1, Roosevelt Rd, Sec. 4 Taipei 10617, Taiwan, R.O.C.; rrao@asiaa.sinica.edu.tw}

\begin{abstract}

We present arcsecond-resolution Submillimeter Array (SMA) polarimetric 
observations of the 880\,$\mu$m continuum emission from the protoplanetary 
disks around two nearby stars, HD 163296 and TW Hydrae.  Although previous 
observations and theoretical work have suggested that a 2-3\% polarization 
fraction should be common for the millimeter continuum emission from such 
disks, we detect no polarized continuum emission above a 3$\sigma$ upper limit 
of 7\,mJy in each arcsecond-scale beam, or $<1$\% in integrated continuum 
emission.  We compare the SMA upper limits with the predictions from the 
exploratory \citet{cho07} model of polarized emission from T Tauri disks 
threaded by toroidal magnetic fields, and rule out their fiducial model at 
the $\sim10\sigma$ level.  We explore some potential causes for this 
discrepancy, focusing on model parameters that describe the shape, magnetic 
field alignment, and size distribution of grains in the disk.  We also 
investigate related effects like the magnetic field strength and geometry, 
scattering off of large grains, and the efficiency of grain alignment, 
including recent advances in grain alignment theory, which are not considered 
in the fiducial model.  We discuss the impact each parameter would have on the 
data and determine that the suppression of polarized emission plausibly arises 
from rounding of large grains, reduced efficiency of grain alignment with 
the magnetic field, and/or some degree of magnetic field tangling (perhaps 
due to turbulence).  A poloidal magnetic field geometry could also reduce the
polarization signal, particularly for a face-on viewing geometry like the 
TW~Hya disk.  The data provided here offer the most stringent limits to date 
on the polarized millimeter-wavelength emission from disks around young stars. 

\end{abstract}


\section{Introduction}
\setcounter{footnote}{0}

The magnetic properties of circumstellar disks are central to a wide range 
of physical processes relevant for planet formation.  Dust and gas 
transport and mixing \citep[e.g.][]{cie07}, meteoritic composition 
\citep[e.g.][]{bos04}, disk chemistry \citep[e.g.][]{sem06}, and the migration 
of planetary embryos through the disk \citep[e.g.][]{cha06} are all thought 
to be influenced by magnetohydrodynamic (MHD) turbulence.  But perhaps the 
greatest impact of a magnetized disk is that MHD turbulence can provide the 
source of viscosity that drives disk evolution.  Since the seminal work by 
\citet{lyn74}, the photospheric excess and variability exhibited by pre-main 
sequence stars have been attributed to an accretion disk.  The viscous 
transport mechanism that supports the accretion process can also explain 
many aspects of the time evolution of circumstellar disks \citep{har98}, and 
by extension can help to constrain the physical conditions and timescales 
relevant for planet formation.  However, there are remarkably few observational 
constraints on the magnitude and physical origin of viscosity in circumstellar 
disks.  

As conjectured by \citet{sha73}, turbulence can provide large enough 
viscosities to account for accretion and disk evolution on the appropriate 
timescales.  The mechanism most commonly invoked as the source of this 
turbulence is the magnetorotational instability (MRI), in which magnetic 
interactions between fluid elements in the disk combine with an outwardly 
decreasing velocity field to produce torques that transfer angular momentum 
from the inner disk outwards (Balbus \& Hawley 1991, 1998; see also Velikhov 
1959 and Chandrasekhar 1960)\nocite{bal91,bal98,vel59,cha60}.  Indeed, it is 
unlikely that turbulence in an unmagnetized, azimuthally symmetric Keplerian 
disk can sufficiently redistribute angular momentum: magnetic fields 
must be invoked to enable Shakura-Sunyaev viscosity \citep[e.g.][]{bal96}.  
The ionization fraction is likely high enough for magnetic coupling of 
material over much of the outer disk \citep[see e.g.][]{san00,tur07}, and 
the observed Keplerian rotation of protoplanetary disks provides the requisite 
velocity shear. However, the magnetic field properties (strength and geometry) 
far from the central star remain unconstrained.

Resolved observations of polarized submillimeter continuum emission are
uniquely suited to constrain the magnetic field geometry -- independent of disk
structure -- via the orientation of polarization vectors produced by dust 
grains aligned with the magnetic field \citep{ait02}.  
In the presence of an anisotropic radiation field, irregularly shaped 
grains with different cross sections to left and right circular polarizations 
of light can be spun up to high speeds by radiative torques 
\citep[e.g.][]{dol72,dol76,dra96}\footnote{More recent research in 
\citet{laz07b} shows that in many cases rather than being spun up paramagnetic 
grains get slowed down by radiative torques, which means that the grains 
aligned by radiative torques do not necessarily rotate suprathermally.  
Nevertheless, the maximal rotational rate provides a useful parameterization 
of the effect of the radiative torques as discussed in detail in 
\citet{hoa08}.}.
These spinning grains precess around magnetic field lines, and 
ultimately align with their long axes perpendicular to the local magnetic 
field direction.  Polarized emission or absorption by these aligned 
grains can thus trace magnetic field structure in dusty interstellar media 
\citep[see][and references therein]{laz07}.  

The first models of polarized emission from disks incorporating the radiative
torque alignment mechanism have recently been calculated by \citet{cho07}.  
Using a two-layer \citet{chi01} disk structure model threaded by a toroidal 
magnetic field (with circular field lines in the plane of the disk, centered 
on the star), they calculated the polarization emitted as a function of
wavelength and position in the disk, incorporating emission and selective 
absorption mechanisms, but not scattering.
They predict a 2-3\% polarization fraction at 850\,$\mu$m, and note that
grain alignment is particularly efficient in the low-density outer disk 
regions.  At millimeter wavelengths, dust grain opacities are low and 
optically thin thermal continuum emission primarily originates in the 
midplane where most of the mass is located.  Polarimetric observations of 
millimeter-wavelength dust continuum emission therefore trace magnetic field 
geometry near the midplane in the outer disk, in regions where the magnetic 
field is strong enough for grains to become aligned and the density is low 
enough that grain spin-up is not impeded by gas drag.

The first attempt to observe polarized millimeter-wavelength emission from
protoplanetary disks was made by \citet{tam95,tam99}.  They used the James 
Clerk Maxwell Telescope (JCMT) to observe several young systems in the 
Taurus-Auriga molecular cloud complex -- HL Tau, GG Tau, DG Tau and GM Aur -- 
and reported tentative ($\sim 3\sigma$) detections of polarized 
millimeter-wavelength continuum emission from three of the four systems.  
The exception was GG Tau, for which they report a 2$\sigma$ upper limit of 
3\%.  While the disks are unresolved in the 14'' JCMT beam, the approximate 
alignment of the GM Aur and DG Tau polarization vectors with the known 
orientation of the disk minor axis is suggestive of a globally toroidal 
magnetic field structure.  However, finer resolution is required to confirm 
the magnetic field structure in the disks, and differentiate it from any 
potential contamination from an envelope or cloud material.  DG Tau was 
followed up at a wavelength of 350\,$\mu$m using the Caltech Submillimeter 
Observatory by \citet{kre09}, and no polarization was detected with an upper 
limit of $\sim 1$\%.  They suggest that the decrease of polarization 
percentage relative to the tentative 850\,$\mu$m detection and the 
corresponding \citet{cho07} prediction at 350\,$\mu$m may be due to some 
combination of polarization self-suppression -- effectively an absorption 
optical depth effect \citep[see e.g.][]{hil00} -- or increased scattering 
at shorter wavelengths, which would produce a signal orthogonal to that 
expected for a toroidal magnetic field.  In summary, while the \citet{cho07} 
predictions are consistent with the magnitude of the tentative JCMT 
detections, as discussed by \citet{kre09}, the predicted polarization spectrum 
is inconsistent with the 350 and 850\,$\mu$m observations of DG Tau.  

In the absence of spatially resolved observations, the origin of the polarized 
emission in these systems remains unclear.  While the position angle of the
polarized emission observed in the three systems in the Taurus-Auriga complex
suggests association with the circumstellar disk, at least two of these 
sources (DG Tau and HL Tau) are flat-spectrum sources host to jets and likely 
retain envelope material that could aid in generating a polarization signal 
\citep[e.g.]{kit96,dal97}.  \citet{tam95} suggest that the emission from 
HL Tau may arise from an interface region between the disk and a small 
envelope, and that the upper limit for GG Tau may be due to the lack of an 
envelope combined with weak, compact emission from the circumbinary ring.  
Observations of the GM Aur system using the NICMOS instrument on the Hubble 
Space Telescope indicate that it too may host a tenuous remnant outflow and 
envelope \citep{sch03}.  
Nevertheless, with both theoretical predictions and observational evidence 
pointing to a 2-3\% polarization fraction at 850\,$\mu$m in several T Tauri 
disks, resolved observations revealing the magnetic field geometry should 
be possible with current millimeter interferometers for bright disks.  
At $\sim1$'' resolution, such data would be well matched to the size scales 
at which the polarization fraction is expected to be the largest in the 
context of the \citet{cho07} models.

In order to test the \citet{cho07} model predictions and constrain magnetic 
field strengths and geometries, we observed two nearby systems, HD~163296 and 
TW~Hya, with the Submillimeter Array (SMA) polarimeter\footnote{The 
Submillimeter Array is a joint project between the Smithsonian Astrophysical 
Observatory and the Academia Sinica Institute of Astronomy and Astrophysics 
and is funded by the Smithsonian Institution and the Academia Sinica.}.  These 
targets were selected primarily for their large millimeter-wave fluxes to 
maximize the expected polarization signal.  Unlike the previously-observed
Taurus targets, they are isolated from molecular cloud material.  HD~163296 
has a total flux of 1.92\,Jy at 850\,$\mu$m \citep{man94}, while TW Hya has 
a flux of 1.45\,Jy at 800\,$\mu$m \citep{wei89}, predicting a total polarized 
flux of $\sim$40\,mJy in each system.  Polarization of this magnitude should 
be observable with the SMA even if resolved across a few beams.  HD~163296 
is a Herbig Ae star with a mass of 2.3\,M$_\sun$ located at a distance of 
122\,pc \citep{anc98}.  It is surrounded by a flared disk viewed at an 
intermediate inclination of $\sim45^\circ$, observed to extend to at least 
500\,AU in molecular gas and scattered light \citep{ise07,gra00}, which has 
been extensively observed and modeled at millimeter wavelengths \citep{man97,
nat04,ise07}.  TW Hya is a K star located at a distance of only 51\,pc 
\citep{mam05,hof98}.  It hosts a massive circumstellar disk viewed nearly 
face-on at an inclination of 7$^\circ$ and extending to a radius of 
$\sim$200\,AU in molecular gas and scattered light \citep{qi04,rob05}.  
It is also a prototypical example of the class of disks with infrared 
deficits in their spectral energy distribution, known as ``transition'' 
disks. It has been shown to have a central deficit of dust emission 
extending out to 4\,AU \citep{cal02,hug07}, with a low mass accretion rate 
\citep{muz00} that may indicate clearing by a giant planet in formation 
\citep{ale07}.  

We describe our observations of these systems in Section~\ref{sec:obs} and
present the upper limits in Section~\ref{sec:res}.  In Section~\ref{sec:model}
we describe the initial predictions generated by the \citet{cho07} models 
and compare these predictions to the SMA observations.  We then use these
initial models as a starting point for an exploration of parameter space that
seeks to describe how the different factors affect the predicted polarization 
properties (Section~\ref{sec:params}).  We expand on these results by 
discussing the potential effects of physical mechanisms not included in the 
models (Section~\ref{sec:other}).  In Section~\ref{sec:sum}, we evaluate which 
physical conditions are most likely to contribute to the suppression of 
polarization relative to the fiducial model and summarize our results. 

\section{Observations and Data Reduction}
\label{sec:obs}

Observations were conducted using the SMA polarimeter, described in detail
in \citet{mar08}.  The polarimeter uses a set of quarter-wave plates to 
convert the normally linear SMA feeds to circular polarization.  By rotating
the wave plate between two orientations separated by 90$^\circ$, a single 
linear feed can be converted into either of the circular bases (left and 
right, or $L$ and $R$).  Since only one polarization ($L$ or $R$) can be 
sampled on any antenna at any given time, full sampling of all four 
polarization states ($LL$, $RR$, $RL$, and $LR$) for all baselines (numbering 
$N(N-1)/2$, where $N$ is the number of antennas) must be accomplished by 
rotating the waveplates through a series of orientation patterns conveniently 
described by the two-state Walsh functions \citep[see][]{mar06}.  A full set 
of polarization states can be obtained for all baselines in $\sim$5\,minutes 
using a series of short (10-second) integrations with the waveplates rotated 
through patterns described by Walsh functions for the appropriate number of 
antennas.  These 5-minute intervals of data are combined into 
quasi-simultaneous Stokes parameters for each baseline ($I = (RR+LL)/2$, 
$Q=(RL+LR)/2$, $U=i(LR-RL)/2$, $V=(RR-LL)/2$), where $I$ measures the total 
intensity, $V$ measures the intensity of circular polarization, and 
$\sqrt{Q^2+U^2}$ gives the total linearly polarized intensity, which can also 
be used to calculate the local fractional (or percent) linear polarization 
$\sqrt{Q^2+U^2}/I$.  Note that the aligned states ($RR$ and $LL$) separate the 
total intensity from the circularly polarized intensity, while the crossed 
states ($LR$ and $RL$) provide information about the linearly polarized 
intensity.  Of course, this represents an idealization. Two relevant non-ideal 
effects are (1) if the $R$ and $L$ gains are not perfectly matched, then some 
of the bright Stokes $I$ flux can leak into Stokes $V$ when the difference is 
taken between $LL$ and $RR$, and (2) instrumental ``leakage'' of left circularly
polarized light through a nominally right circularly polarized waveplate 
(and vice versa) can transfer Stokes $I$ to the linear states $Q$ 
and $U$.  

Polarimetric SMA observations of the HD~163296 disk at 880\,$\mu$m wavelength
were carried out in the compact configuration on 29 May 2008, and in the
extended configuration on 12 July 2008.  The weather was excellent, with the
225\,GHz opacity below 0.05 both nights, reaching as low as 0.03 on the night
of 12 July.  The phases were also extremely stable on both nights.  The
projected baseline lengths spanned a range of 9 to 260 k$\lambda$, providing 
a synthesized beam size of 1\farcs1$\times$0\farcs89 for the combined
data set, using natural weighting (see Table~\ref{tab:obs} for details).  The 
quasar 3c454.3 was observed for 2.5 hours through more than 90$^\circ$ of 
parallactic angle during its transit, in order to calibrate the complex 
leakages of the quarter-wave plates.  The quasar J1733-130 was used to 
calibrate the atmospheric and instrumental gain, and the quasar J1924-292 
was observed at 45-minute 
intervals throughout the night to test the quality of the phase transfer from J1733-130 as well as the calibration of the 
quarter-wave plate leakage.  Uranus was used as the flux calibrator, 
yielding a flux for J1733-130 of 1.62 Jy on the night of May 29 and 2.01 Jy 
on the night of July 12.  Uranus, Callisto, 3c273, and 3c279 were included 
as passband calibrators.

Observations of the disk around TW~Hya were conducted in the subcompact
and extended configurations of the SMA during the nights of 25 January and 15 
February 2009, respectively.  Due to the far southern declination of TW~Hya
in combination with the stringent elevation limits imposed to avoid antenna
collisions in the subcompact configuration, the source was only observable for
three hours on the night of 25 January, while a full six hours of observations 
were obtained on 15 February.  The weather was again excellent, particularly 
for the extended configuration, during which the 225\,GHz opacity remained 
stable between 0.03 and 0.04 for most of the night.  The projected baseline 
lengths in the final data set varied from 6 to 250\,k$\lambda$, providing a 
synthesized beam size of 1\farcs2$\times$0\farcs9 in the final data set (see 
Table~\ref{tab:obs}).  The instrumental polarization was calibrated by 
observing 3c273 over 90 degrees of parallactic angle for three hours across 
its transit.  The quasar J1037-295 was used as the gain calibrator, and 
3c279 was observed once per hour to test both the quality of the phase 
transfer and the instrumental polarization calibration.  The primary flux 
calibrator was Titan, yielding a flux for J1037-295 of 0.64\,Jy on the night 
of 25 Jan and 0.53\,Jy on the night of 15 Feb.  3c279, 3c273, and J1037-295 
were included as passband calibrators. 

The double sideband receivers were tuned to a central frequency of 340.75\,GHz
(880 $\mu$m) for the HD~163296 observations and 341.44\,GHz (877\,$\mu$m) for 
the TW~Hya observations, with each 2\,GHz-wide sideband centered $\pm$5\,GHz
from that value.  The correlator was configured to observe the CO(3-2) 
transition (rest frequency 345.796\,GHz) with a velocity resolution of 
0.70\,km\,s$^{-1}$.  The data were edited and calibrated using the MIR 
software package\footnote{See http://cfa-www.harvard.edu/$\sim$cqi/mircook.html.}, while the standard tasks of Fourier transforming the visibilities, 
deconvolution with the CLEAN algorithm, and restoration were carried out
using the MIRIAD software package.  For a summary of the observational 
parameters, including the 3$\sigma$ upper limits in Stokes $Q$ and $U$ for the 
individual tracks and the combined data sets, refer to Table~\ref{tab:obs}.  
The test quasars for all tracks were point-like and unresolved. We detect 
polarized emission from the test quasars independently in each data set 
with a polarization fraction of between 8 and 12\% and a direction consistent
between lower and upper sidebands, as expected for linearly polarized emission 
from quasars at these wavelengths \citep[see][]{mar06}.  

\begin{table*}[ht]
\caption{Observational Parameters$^a$}
\resizebox{\textwidth}{!}{ 
\begin{tabular}{|l|ccc|ccc|}
\hline
 & \multicolumn{3}{c|}{HD~163296} & \multicolumn{3}{c|}{TW~Hya} \\
\cline{2-7}
 & Compact & Extended & C+E & Compact & Extended & C+E \\
Parameter & 29 May 2008 & 12 July 2008 & & 25 Jan 2009 & 15 Feb 2009 & \\
\hline
\multicolumn{7}{|c|}{340\,GHz Continuum} \\
\hline
Beam Size (FWHM) & 2\farcs2$\times$1\farcs3  & 0\farcs9$\times$0\farcs7 & 1\farcs0$\times$0\farcs9 & 4\farcs7$\times$2\farcs0 & 1\farcs1$\times$0\farcs8 & 1\farcs2$\times$0\farcs8\\
~~~P.A. & 50$^\circ$ & -8$^\circ$ & 7$^\circ$ & -1$^\circ$ & 6$^\circ$ & 6$^\circ$ \\
RMS Noise (mJy\,beam$^{-1}$) & \multicolumn{3}{c|}{ } & \multicolumn{3}{c|}{ } \\
~~~Stokes $I^b$ & 11 & 7.8 & 5.7 & 35 & 6.8 & 5.5 \\
~~~Stokes $Q$ \& $U$ & 3.8 & 2.7 & 2.4 & 6.3 & 2.4 & 2.3 \\
Peak Flux Density (mJy\,beam$^{-1}$) & \multicolumn{3}{c|}{ } & \multicolumn{3}{c|}{ } \\
~~~Stokes $I$ & 996 & 639 & 739 & 990 & 450 & 474 \\
~~~Stokes $Q$ \& $U$ (3$\sigma$ upper limit) & $<$11 & $<$8.1 & $<$7.2 & $<$19 & $<$7.2 & $<$6.9 \\
Integrated Flux$^c$ (Stokes $I$; Jy) & 1.65 & 1.79 & 1.64 & 1.24 & 1.33 & 1.26 \\
\hline
\multicolumn{7}{|c|}{CO(3-2) Line} \\
\hline
Beam Size (FWHM) & 2\farcs2$\times$1\farcs4 & 0\farcs9$\times$0\farcs7 & 1\farcs1$\times$1\farcs0 & 5\farcs0$\times$1\farcs2 & 1\farcs1$\times$0\farcs7 & 1\farcs2$\times$0\farcs7 \\
~~~P.A. & 50$^\circ$ & -8$^\circ$ & 17$^\circ$ & -1$^\circ$ & 9$^\circ$ & 9$^\circ$ \\
RMS Noise (mJy\,beam$^{-1}$) & 25 & 19 & 15 & 55 & 14 & 13 \\
Peak Flux Density (mJy\,beam$^{-1}$) & \multicolumn{3}{c|}{ } & \multicolumn{3}{c|}{ } \\
~~~Stokes $I$ & 6500 & 2650 & 3730 & 1520 & 1800 & 3090 \\
~~~Stokes $Q$ \& $U^d$ (3$\sigma$ upper limit) & $<$75 & $<$57 & $<$45 & $<$170 & $<$42 & $<$39 \\
Integrated Flux$^e$ (Stokes $I$; Jy\,km\,s$^{-1}$) & 110 & 56 & 95 & 47 & 13 & 27 \\
\hline
\end{tabular}
}
\tablenotetext{a}{All quoted values assume natural weighting.}
\tablenotetext{b}{The rms in Stokes I is limited by dynamic range rather than
sensitivity.}
\tablenotetext{c}{The integrated continuum flux is calculated using the 
MIRIAD task \texttt{uvfit}, assuming an elliptical Gaussian 
brightness profile. }
\tablenotetext{d}{The rms for the line is calculated using a channel width of 
0.7\,km\,s$^{-1}$.}
\tablenotetext{e}{The integrated line flux is calculated by integrating the
zeroeth moment map inside the 3$\sigma$ brightness contours.}
\label{tab:obs}
\end{table*}

\section{Results}
\label{sec:res}

\begin{figure*}[ht]
\epsscale{1.0}
\plotone{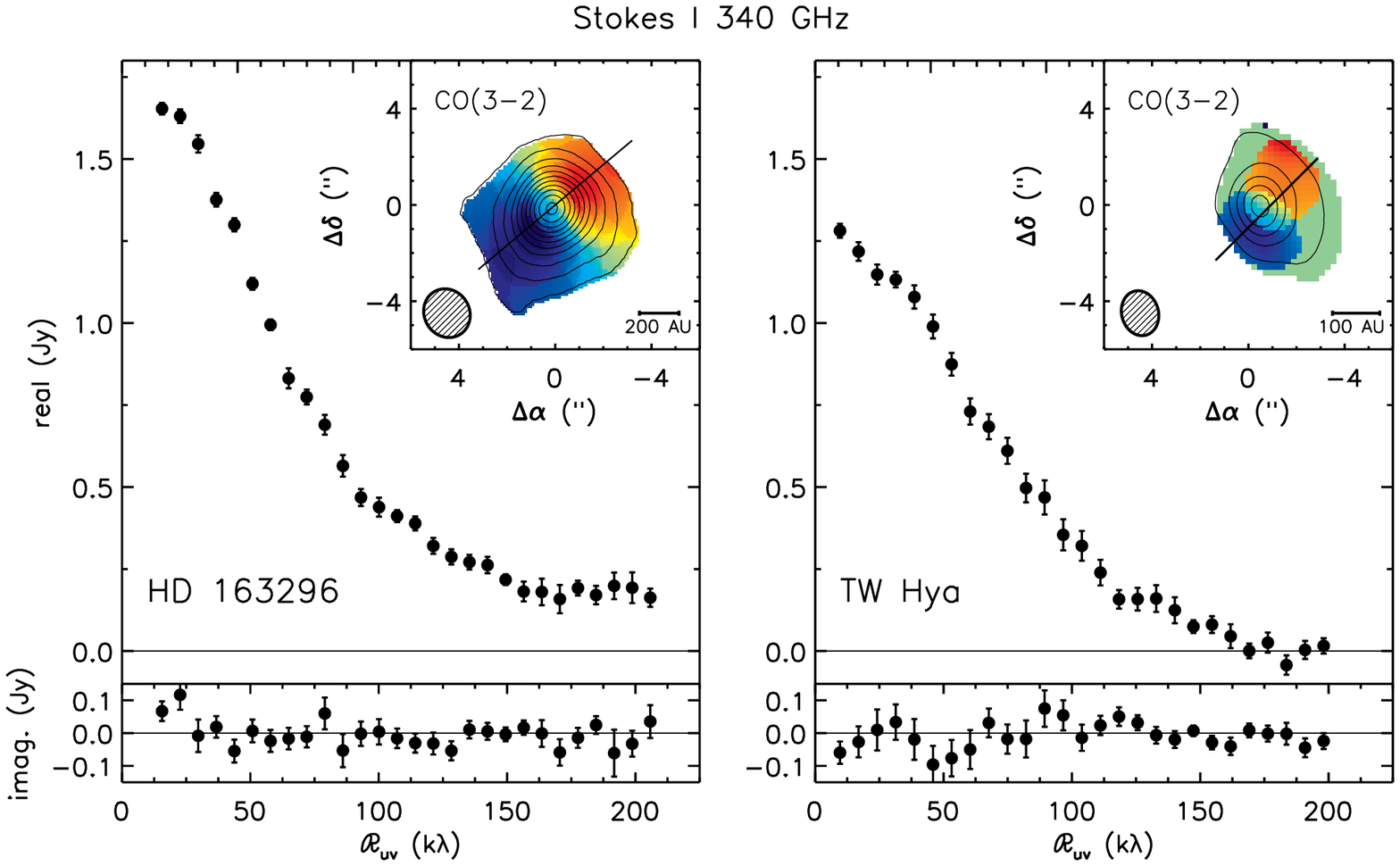}
\figcaption{
Real (top) and imaginary (bottom) Stokes $I$ continuum visibilities for HD~163296 
(left) and TW~Hya (right) as a function of distance from the disk center in the 
($u$,$v$) plane, corrected for projection effects due to the inclination of 
the disk to our line of sight.  Error bars show the standard error of the 
mean in each 7\,k$\lambda$ bin.  See \citet{lay97} for details of the 
deprojection process.  The inset in the upper right of each plot shows the
CO(3-2) moment maps in Stokes $I$ for the two disks.  The colors indicate
the first moment (intensity-weighted velocity), and the contours show the
zeroeth moment (velocity-integrated intensity) in intervals of 
3\,Jy\,km\,s$^{-1}$. The solid line marks the position angle of the disk
as determined by \citet{ise07} and \citet{qi04}.  The size and orientation
of the synthesized beam is indicated at the lower left of each moment map.
\label{fig:vis}}
\end{figure*}

Figure~\ref{fig:vis} shows the Stokes $I$ (unpolarized) visibilities as a 
function of distance from the phase center in the ($u$,$v$) plane, corrected 
for the projection effects due to the disk inclination as in \citet{lay97} 
\citep[for the mathematical definition of the abscissa, see Section 3.3 
of ][]{hug08}.  This is effectively the Fourier transform of the radial 
brightness distribution of the disk.  Both the HD~163296 and TW~Hya disks are 
well resolved with high signal-to-noise ratios.  

\begin{figure*}[t]
\epsscale{1.0}
\plotone{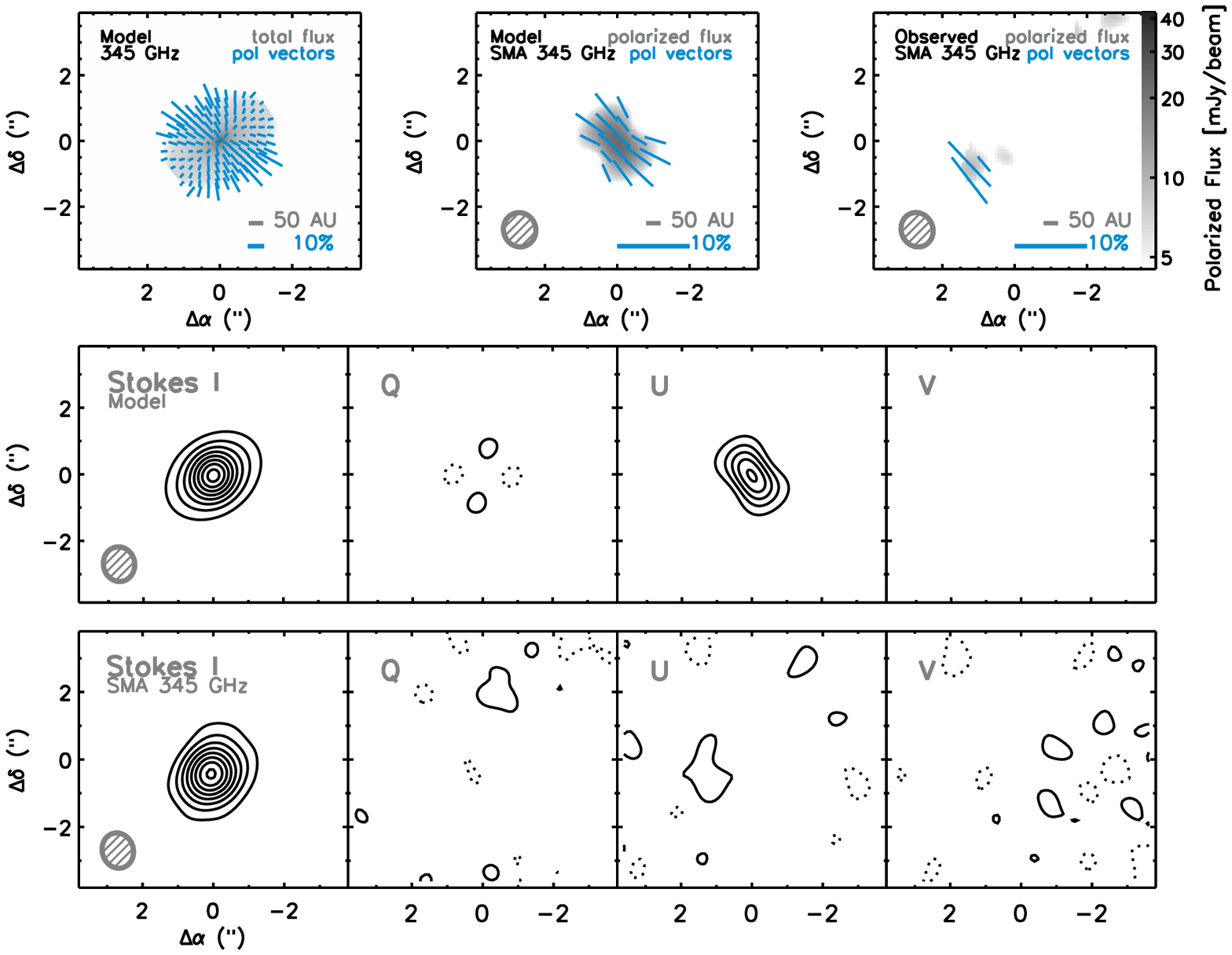}
\figcaption{
Comparison between the \citet{cho07} model and the SMA 340\,GHz observations
of HD~163296.  The top row shows the prediction for the
model at full resolution (left), a simulated observation of the model
with the SMA (center), and the 2008 SMA observations (right).  The grayscale 
shows either the total flux (left) or the polarized flux (center, right), 
and the blue vectors indicate the percentage and direction of polarized flux 
at half-beam intervals.  The center and bottom rows compare the model 
prediction (center) with the observed SMA data (bottom) in each of the four 
Stokes parameters ($I$, $Q$, $U$, $V$, from left to right).  Contour levels 
are the same in both rows, either multiples of 10\% of the peak flux (0.9 
Jy/beam) in Stokes $I$ or in increments of 2$\sigma$ for $Q$, $U$, and $V$, 
where $\sigma$ is the rms noise of 2.4\,mJy/beam.  The size and orientation 
of the synthesized beam is indicated in the lower left of each panel.
\label{fig:hd163296}
}
\end{figure*}

We detect no polarized emission, in the CO(3-2) line or 880\,$\mu$m continuum, 
from the HD~163296 or TW~Hya disks.  The rms values achieved in Stokes $Q$ 
and $U$ for the combined (compact+extended) continuum data are 
2.4\,mJy\,beam$^{-1}$ and 
2.3\,mJy\,beam$^{-1}$, respectively, yielding a 3$\sigma$ upper limit in 
both data sets of 7\,mJy\,beam$^{-1}$.  Given the integrated Stokes $I$ fluxes 
of 1.65\,Jy and 1.25\,Jy for HD~163296 and TW~Hya (see Table~\ref{tab:obs}), 
the \citet{cho07} result predictions of 2-3\% polarization at these 
wavelengths imply $\sim$30-50\,mJy of polarized flux.  Even if the spatial 
distribution of polarized flux in the source differs from that of the 
unpolarized emission, we should be able to detect it given that we recover 
most of the Stokes I flux.  Figures~\ref{fig:hd163296} and \ref{fig:twhya} 
compare the data with the fiducial model predictions (described in 
Section~\ref{sec:model} below).  The upper right panel of each figure 
displays the amount and direction of observed polarized flux for each 
source, while the bottom row presents contour maps for each of the individual 
Stokes parameters.  The emission in Stokes $Q$ and $U$ (linear polarization), 
as well as in Stokes $V$ (circular polarization), is consistent with noise.  
As noted in Section \ref{sec:obs}, since Stokes $V$ is calculated as the 
difference between the measured right and left ($RR$ and $LL$) circular 
polarization, the difficulty of calibrating the gains precisely enough to 
remove the influence of the bright Stokes $I$ emission raises the rms value 
in this Stokes parameter relative to Stokes $Q$ and $U$, which are calculated 
instead from the crossed ($RL$ and $LR$) polarization states.

We can rule out calibration errors as the reason for the lack of polarized
emission for three reasons: (1) The point-like test quasars and the similarity 
of the visibility profiles in Figure~\ref{fig:vis} with previous observations
of these sources \citep[see e.g.][]{ise07,hug08} illustrate both the success 
of the atmospheric and instrumental gain calibration and the high sensitivity 
of the data set.  (2) The detection of polarized emission from the test quasars
in each of the data sets, with direction consistent between sidebands, 
demonstrates the success of the instrumental leakage calibration.  
Furthermore, (3) several of the nights were shared with other SMA polarization 
projects and our solutions for the instrumental leakage between Stokes 
parameters for the eight quarter-wave plates were effectively identical to 
those derived by other observers, who successfully detect polarization in 
their targets.

It is worthwhile to compare the rms noise achieved here with the
limiting precision of the current SMA polarimeter.  Errors in alignment of
the quarter-wave plates introduce instrumental ``leakage'' between 
Stokes parameters, allowing some of the flux from Stokes $I$ to bleed into
the linear Stokes parameters.  The instrumental leakage correction is quite 
small ($\lesssim 3$\%) and can to a large extent be calibrated by observing 
a bright point source as it rotates through 90$^\circ$ of parallactic angle.  
Nevertheless, the uncertainty of this correction under typical observing 
conditions is $\sim$0.2\%, although this can be reduced to $\lesssim0.1$\% 
with parallactic angle rotation, provided the source polarization does not 
vary with time \citep{mar06}.  Given the 2\,mJy\,beam$^{-1}$ rms noise from 
our observations compared with the peak Stokes $I$ fluxes of 740 and 
470\,mJy\,beam$^{-1}$ ($\sim$0.3\%), our constraints on the polarized flux 
are approaching the limit of what is achievable with the SMA polarimeter. 

It is difficult to directly compare the observations presented here with
the \citet{cho07} model predictions and the \citet{tam99} JCMT result.  The 
2-3\% polarization factor reported by both sources refers to the integrated 
emission over the entire spatial extent of the disk.  Since the SMA spatially 
resolves the emission from the disk, the limit on the percent polarization 
varies with position across the disk. The emission structure is predicted to 
be quite complicated \citep{cho07}, with the percent polarization increasing as
a function of distance from the star, so there is no straightforward way to 
quote a single value for the percent polarization that can be easily compared 
with the data.  By tapering the SMA visibilities with a Gaussian whose FWHM 
is equal to the diameter of the disk as measured by a truncated power law 
model \citep{hug08}, we can simulate an unresolved observation, similar to 
the JCMT result from \citet{tam99}.  Using this method, we place a 3$\sigma$ 
upper limit of 1\% on the total polarized flux from both disks.  However, 
such an extreme taper severely down-weights the visibilities on the longest 
baselines, which still have very high signal-to-noise ratios (see 
Figure~\ref{fig:vis}).  This effectively neglects the majority of the data: 
when all of the spatially resolved data are taken into account, the limits 
are much more stringent, but they must be compared with the more complicated 
predictions from the spatially resolved model.  Furthermore, decreasing the 
resolution may be additionally detrimental in the case of more face-on disks 
like TW Hya: if the magnetic field is perfectly toroidal, then the resulting 
radial polarization signal will cancel to zero in a large beam, no matter 
how strong the emission.  To give a rough estimate, the $\sim40$\,mJy of 
integrated polarized flux predicted for a 2-3\% polarization fraction resolved 
into a few beams might predict a peak flux density of $\sim20$\,mJy\,beam$
^{-1}$, which is about 10$\sigma$ above the $\sim2$\,mJy\,beam$^{-1}$ noise 
in the data.  However, a detailed comparison with the spatially resolved
model predictions for each disk can give a more robust result. 

The highest signal-to-noise ratio in an image is achieved using natural 
weighting, which assigns each visibility a weight inversely proportional to 
its variance.  In the case of observations with the SMA polarimeter, the 
bandwidth and integration time are the same for each integration, so the 
visibilities are primarily weighted by system temperature.  For this reason, 
we use natural weighting to generate all images presented here.  Using the 
upper limits from the naturally weighted images, it is possible to make 
comparisons with predictions of the spatially resolved emission generated 
from the models of \citet{cho07}.  We pursue this avenue of investigation 
in the following section. 

\begin{figure*}[t]
\epsscale{1.0}
\plotone{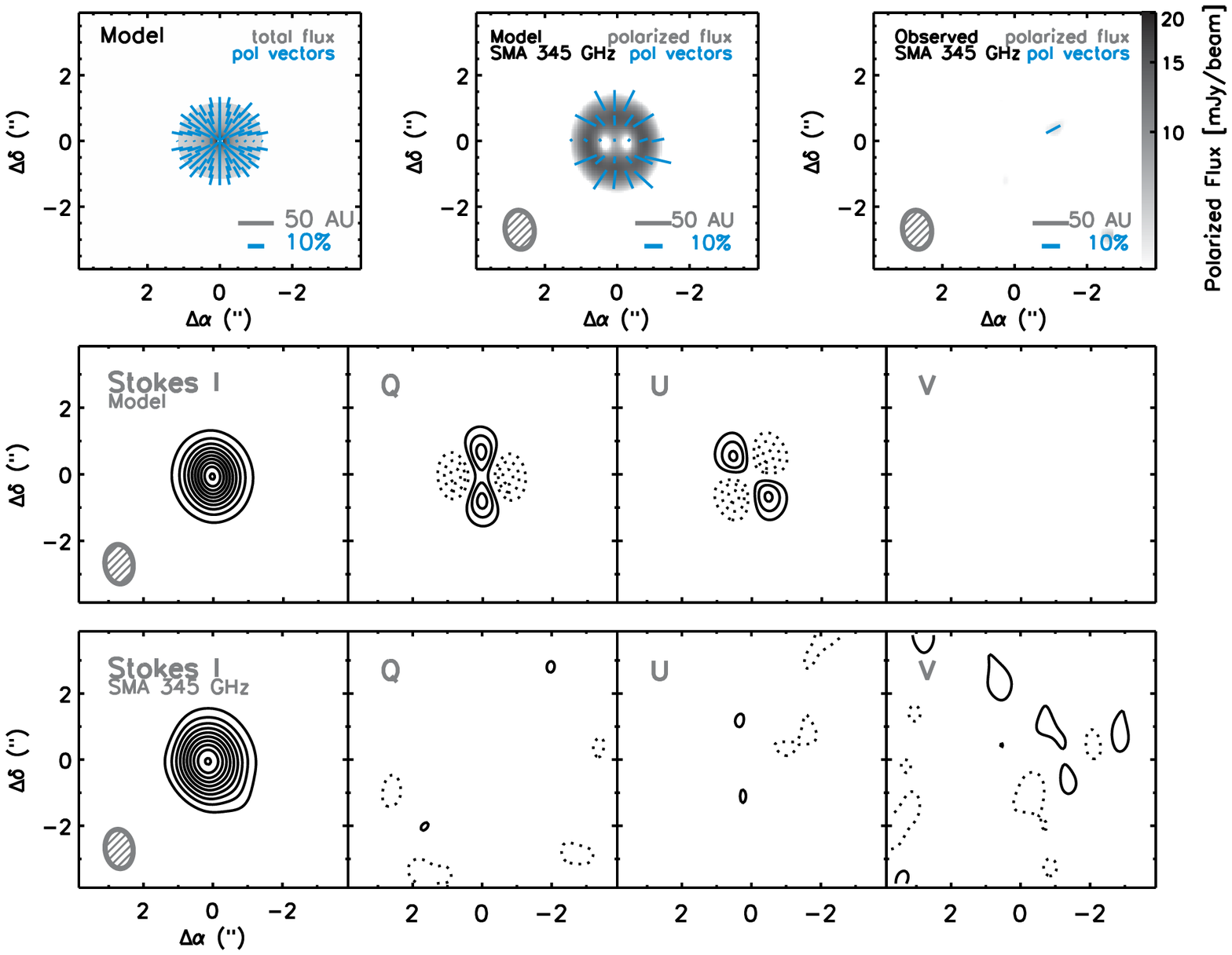}
\figcaption{
Comparison between the \citet{cho07} model and the SMA 340\,GHz observations
of TW~Hya.  The top row shows the prediction for the model at full resolution 
(left), a simulated observation of the model with the SMA (center), and the 
SMA observations (right).  The center and bottom rows compare the model
prediction (center) with the observed SMA data (bottom) in each of the four
Stokes parameters ($I$, $Q$, $U$, $V$, from left to right).  Contour levels 
are the same in both rows, either multiples of 10\% of the peak flux 
(47\,mJy/beam) in Stokes $I$ or at 2$\sigma$ intervals for $Q$, $U$, and $V$, 
where $\sigma$ is the rms noise of 2.3\,mJy/beam.  Symbols as in 
Figure~\ref{fig:hd163296}.
\label{fig:twhya}}
\end{figure*}

\section{Analysis and Discussion}
\label{sec:analysis}

The constraints on polarized millimeter wavelength emission from the disks
around TW~Hya and HD~163296 are inconsistent with previous observational 
\citep{tam99} and theoretical \citep{cho07} work that suggested that a 
polarization fraction of 2-3\% should be common among protoplanetary disks.  
The stringent limit on the polarization fraction, when investigated within the 
context of the \citet{cho07} model, can provide clues to the physical 
conditions within the disk that may be responsible for the suppression of 
polarized emission relative to the fiducial model prediction.  We therefore 
use the code described in \citet{cho07} to generate models of the emission 
predicted for the TW~Hya and HD~163296 disks, using available observational
constraints on the disk properties as inputs, and compare these predictions 
to the upper limits from the SMA observations (Section~\ref{sec:model}).  We 
then identify parameters that are not well constrained by existing 
observations, and which have the greatest effect on the polarized emission 
rather than unpolarized Stokes $I$ emission.  We vary these parameters and 
investigate their effects on the predicted polarized submillimeter emission.  
We infer the range of values over which the predictions are consistent with 
the observations as well as the interactions between parameters in the context 
of the models (Section~\ref{sec:params}).  Finally, we investigate other 
effects {\it not} implemented in these models that may contribute to the 
suppression of polarized disk emission, and estimate the magnitude of their 
contribution (Section~\ref{sec:other}).  

\subsection{Initial Models}
\label{sec:model}

The \citet{cho07} predictions employ a two-layered \citet{chi01} model of the
density and temperature structure of a protoplanetary disk, including a 
surface layer with hot, small dust grains and an interior with cooler, larger
grains.  Within this model, the elongated dust grains are allowed to align 
via the radiative torque mechanism with a perfectly toroidal magnetic field 
threading the disk.  The dust grains are assigned a size distribution 
described by a power law $dN \propto r^{-q_\mathrm{grain}} dr$ where $N$ 
is the number of grains of size $r$, and $q_\mathrm{grain}$ is initially 
taken to be 3.5 \citep{mat77}.  
The grains are also assigned a degree of elongation given by the ratio of 
long-to-short axis cross sections, $C_\perp / C_\parallel$, where $C_\perp$ and
$C_\parallel$ are the polarization cross sections for the electric field
perpendicular and parallel to the grain symmetry axis, respectively.  
The grain size is defined as $r$, such that $C_\perp = (1+\alpha) \pi r^2$ and
$C_\parallel = (1-\alpha) \pi r^2$, where $\alpha$ parameterizes the degree of
elongation.  The ratio of the major and minor axes of the grain are then
given by $a/b = \sqrt{(1+\alpha)/(1-\alpha)}$.  The grain shape is assumed
to be oblate as in \citet{cho07}, consistent with observational evidence
described in \citet{hil95}.  The initial 2-3\% polarization 
estimates are based on the parameters for the ``typical'' T Tauri disk 
investigated in \citet{chi01}.  

In order to generate a model prediction that can be compared with the
upper limits from the SMA observations, we adjust these parameters to
reflect the best available information about the grain properties and 
density structures in the disks around HD~163296 and TW~Hya.  The initial 
model inputs, with references, are summarized in Table~\ref{tab:model}.  
We use temperature and surface density power law indices and outer radii 
derived from previous SMA 345\,GHz continuum observations \citep{hug08}.  
The temperatures are calculated from the stellar temperature and gas and 
dust densities and opacities as in \citet{chi01}, while the surface density 
is adjusted to best reproduce the observed 880\,$\mu$m continuum flux.  The 
temperatures and surface densities calculated here are consistent 
with previously determined values \citep[e.g.][]{ise07,hug08} to within a 
factor of two.  Variations can be attributed to differences in the vertical 
temperature structure and dust grain opacities assumed in the models.
While these disk structure models do not precisely reproduce the observed 
brightness profile, they represent a reasonable approximation within which 
the parameters determining the polarization properties of interest can 
be investigated.

\begin{table}[ht]
\caption{Initial Model Parameters}
\begin{tabular}{lcccc}
\hline
 & \multicolumn{2}{c}{HD~163296} & \multicolumn{2}{c}{TW~Hya} \\
Parameter$^a$ & Value & Ref. & Value & Ref. \\
\hline
\hline
$T_*$ (K) & 9330 & 1 & 4000 & 2 \\
$R_*$ (R$_\sun$) & 2.1 & 1 & 1.0 & 2 \\
$M_*$ (M$_\sun$) & 2.3 & 1 & 0.6 & 2 \\
$p$ & 0.8 & 3 & 1.0 & 3 \\
$a_\mathrm{inner}$ (AU) & 0.45 & 4 & 4.0 & 5,6 \\
$a_0$ (AU) & 200 & 3 & 60 & 3 \\
$r_\mathrm{max,i}$ ($\mu$m) & $10^3$ & 4 & $10^4$ & 7 \\
$i$ & 46$^\circ$ & 4 & 7$^\circ$ & 8 \\
$d$ (pc) & 122 & 1 & 51 & 9,10 \\
$\Sigma_0$ (g cm$^{-2}$) & 130 & -- & 170 & -- \\
\hline
\end{tabular}
\tablenotetext{a}{Symbols as in \citet{chi01}: $T_*$, $R_*$, and $M_*$ are 
stellar temperature, radius, and mass, respectively; $p$ and $\Sigma_0$ 
describe the surface density profile $\Sigma(R) = \Sigma_0 (R/1\mathrm{AU}
)^{-p}$; $a_0$ is the outer disk radius; and $r_\mathrm{max,i}$ is the 
maximum dust grain size in the disk interior. Additionally, we define 
$a_\mathrm{inner}$ (disk inner radius), $i$ (inclination), and $d$ 
(distance).  All parameters not listed here are equal to the fiducial 
input parameters from \citet{chi01}.}
\tablerefs{~
(1) \citet{anc98};
(2) \citet{web99};
(3) \citet{hug08};
(4) \citet{ise07};
(5) \citet{cal02};
(6) \citet{hug07};
(7) \citet{wil05};
(8) \citet{qi04};
(9) \citet{mam05};
(10) \citet{hof98}
}
\label{tab:model}
\end{table}

We use the model routines to generate 400$\times$600 pixel sky-projected 
images (i.e. with 6- and 8-milliarcsecond pixels for TW Hya and HD 163296,
respectively, significantly more finely spatially sampled than the data) giving 
the total continuum flux, percent polarization, and orientation of polarized 
emission at each position across the disk.  The full-resolution 
model is shown in the upper left panel of Figures~\ref{fig:hd163296} and 
\ref{fig:twhya}, although the lines indicating orientation have been 
vector-averaged in bins of several pixels for clarity of display.  We then 
use the MIRIAD task \texttt{uvmodel} to sample the image with the same 
spatial frequencies as the SMA data.  We invert the visibilities and image 
with natural weighting to create a simulated SMA observation of the disk 
model, shown in the top center panel of Figures~\ref{fig:hd163296} and 
\ref{fig:twhya}.  We also create simulated images in each of the four 
Stokes parameters (center row), since the Stokes parameter images are most 
directly comparable to the upper limits set by the observations.  The model 
images show the distinctive quadrupolar pattern in Stokes $Q$ and $U$ predicted 
by the model for a toroidal magnetic field geometry, due to the radial 
orientation of the polarization vectors.  The 
intermediate inclination of HD~163296 creates an hourglass-shaped bright 
region along the disk minor axis, where the synthesized beam picks up 
emission from the highly polarized regions along the front and back of 
the outer disk, concentrated towards the disk center by the viewing 
geometry.  This predicted morphology echoes the alignment of polarization 
vectors with the minor axes of the disk observed by \citet{tam99}.  With
predicted peak Stokes $Q$ and $U$ fluxes of 23 and 16 mJy\,beam$^{-1}$, these 
initial models of polarized emission are ruled out at the 10$\sigma$ and 
7$\sigma$ level for HD~163296 and TW~Hya, respectively, by the SMA upper 
limits. 

\subsection{Parameter Exploration}
\label{sec:params}

With the fiducial model prediction ruled out at high confidence, we
turn to an exploration of the input parameter space to provide information
about the conditions in the disk that might be responsible for the suppression
of polarized emission.  We first identify several parameters that most 
strongly affect the polarization properties of the disk, without significant 
impact on the Stokes $I$ emission.  In the \citet{cho07} model, the radiative 
torque mechanism that spins up elongated dust grains along magnetic field 
lines is impeded primarily by gas drag in regions of high density.  Since 
we normalize the surface density to reproduce the 880\,$\mu$m flux (for the
assumed opacities and derived temperatures), we cannot vary this quantity.  
However, the degree of elongation of the dust grains, the threshold set 
within the model for grain alignment, and the dust grain size distribution 
are all important factors that affect the polarization properties of the 
disk rather than the Stokes $I$ emission.  These parameters are discussed in 
greater detail in the following sections. 

\subsubsection{Grain Elongation}
\label{sec:qratin}

The elongation of the dust grains is important both for the radiative torque 
and because the differing cross-sections parallel and perpendicular to the 
magnetic field allow the grain to emit polarized continuum emission at 
millimeter wavelengths.  The fiducial model assumes a long-to-short axis
cross-section ratio $C_\perp / C_\parallel = 2.1$, corresponding to an axial 
ratio of 1.5:1 for oblate dust grains \citep[for the relationship between 
cross section and axial ratios for different grain geometries, see 
e.g.][]{pad01}.  Varying this ratio determines the radial extent of the disk 
over which the dust grains are aligned with the magnetic field, as well as 
how much polarized light is emitted from the disk: it effectively changes the 
efficiency of grain alignment and the emission cross-section of the grains.

\begin{figure*}[ht]
\epsscale{1.1}
\plottwo{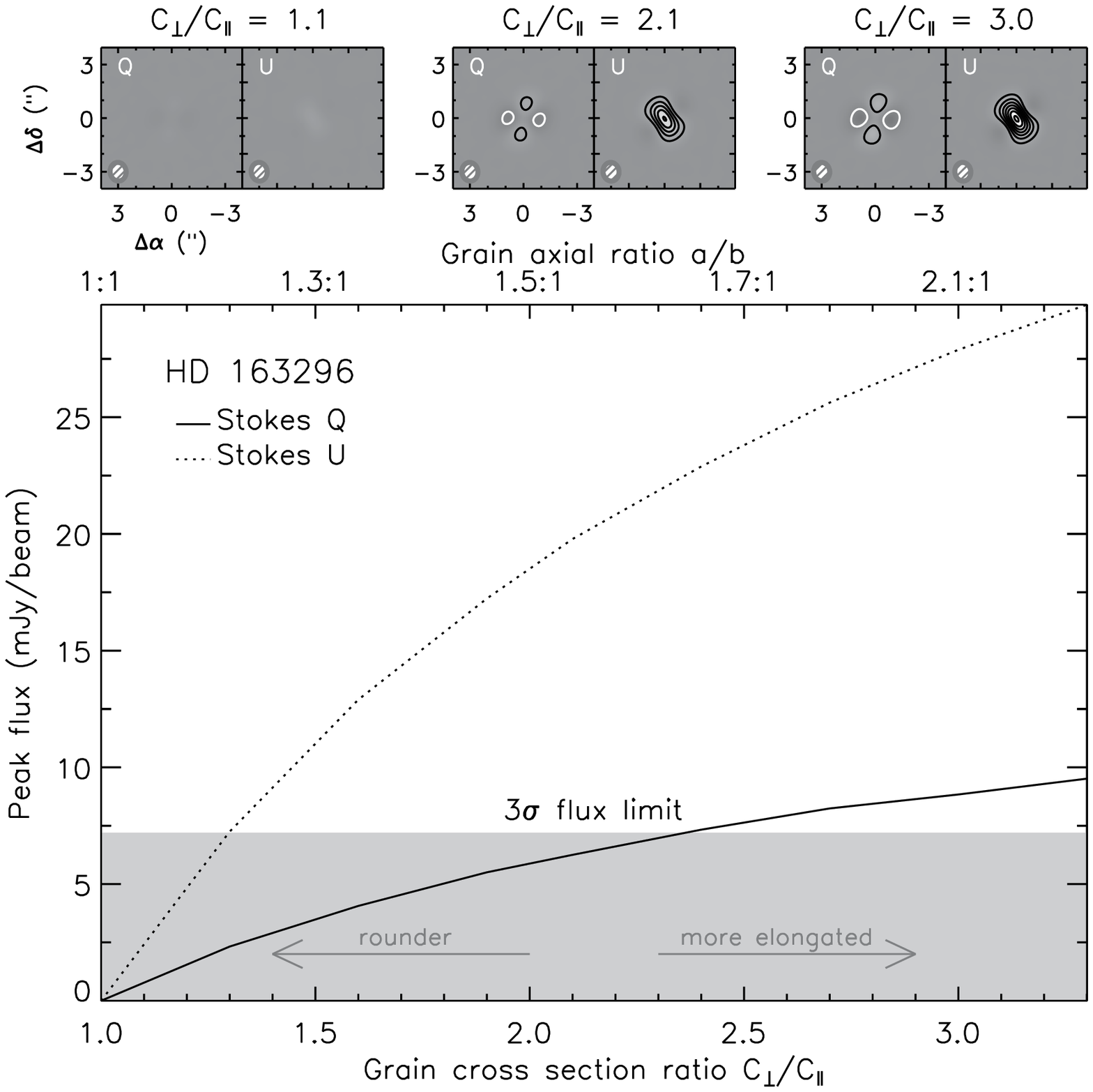}{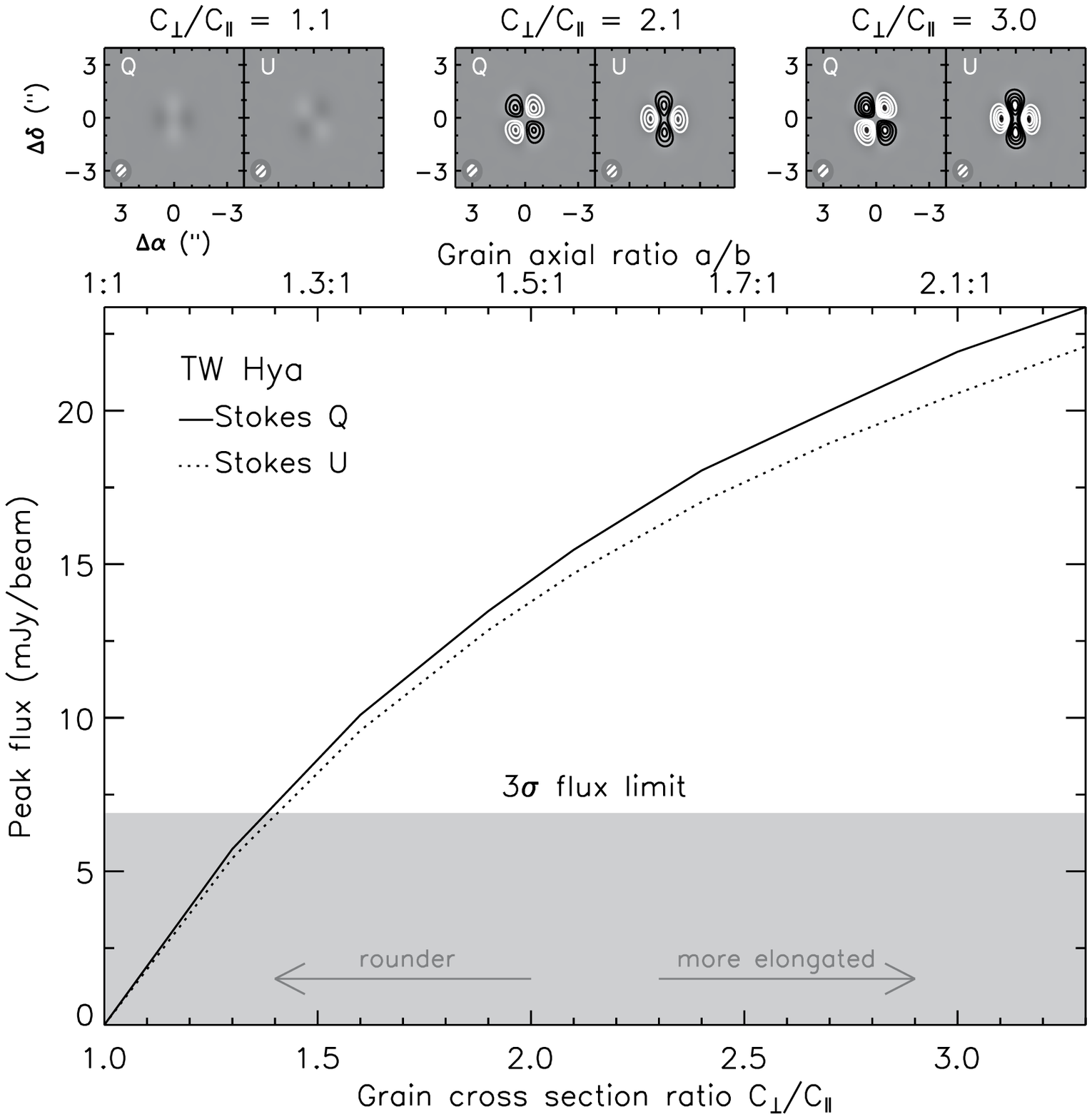}
\figcaption{
Peak continuum flux in Stokes $Q$ and $U$ as a function of dust grain cross 
section ratio for HD~163296 (left) and TW~Hya (right).  The top row 
shows the resolved emission in Stokes $Q$ and $U$ predicted for three values 
of the dust grain cross section ratio, sampled at the same spatial frequencies 
as the data.  The grayscale indicates the intensity of emission relative to 
the peak flux of the data when the grain cross section ratio equals three, 
with white indicating positive emission and black indicating negative 
emission.  Contours are [2,4,6,...] times the rms noise (2.4\,mJy for 
HD~163296 and 2.3\,mJy for TW~Hya) with positive contours in black and 
negative contours in white.  The plots below give the peak flux in the 
synthesized beam predicted by the models as a function of the grain cross 
section ratio.  Stokes $Q$ is plotted as a solid line while Stokes $U$ is a 
dotted line.  The three-sigma upper limit on the peak flux from the SMA 
observations is indicated by the gray region of the plot.  The y-axis along 
the upper edge of the plot gives the dust grain axial ratio.  All images and 
peak flux values assume natural weighting to minimize noise. 
\label{fig:qratin}}
\end{figure*}

In order to obtain a quantitative description of the effect of grain 
elongation on the predicted intensity of polarized emission from the disk,
we generate a series of models with different cross section ratios as 
described in \citet{cho07} with initial parameters listed in Table 
\ref{tab:model}.  We then sample the model images with the SMA spatial 
frequencies, as described in Section~\ref{sec:model} above, and compare 
the peak flux in Stokes $Q$ and $U$ with the 3$\sigma$ upper limit from the 
SMA observations.  Figure~\ref{fig:qratin} plots the peak flux in the 
Stokes $Q$ and $U$ model images as a function of the dust grain cross section 
ratio.  For comparison, the shaded area marks the region of parameter 
space consistent with the 3$\sigma$ upper limits from the SMA observations.  
The series of panels across the top of the plot show the model images in 
Stokes $Q$ and $U$, sampled with the SMA spatial frequencies, for three 
representative values of the dust grain cross section ratio.  From these 
maps, it is clear that the dust grain elongation acts primarily as a 
scaling factor for observations at this resolution: the emission morphology 
does not change, but simply becomes stronger or weaker as the dust grains 
become more or less elongated.  From the HD~163296 plot on the left and 
the TW~Hya plot on the right, we can see that if the dust grain elongation 
were the only factor suppressing polarized emission from the disk, the 
grains would have to be quite round, with $C_\perp / C_\parallel \lesssim 
1.2-1.3$.  

\subsubsection{Grain Alignment Criterion} 
\label{sec:ratioin}

Another model input that is important for the polarization properties of 
the disk is the value at which the threshold for grain alignment via the 
radiative torque is set.  In order to determine whether or not the dust 
grains are aligned with the magnetic field in a particular region of the 
disk, a comparison is made between the rotational kinetic energy imparted 
by the radiative torque and that imparted by random collisions with gas 
particles in the disk.  A useful parameterization is $(\omega_\mathrm{rad} / 
\omega_\mathrm{th})^2$, where $\omega_\mathrm{rad}$ and $\omega_\mathrm{th}$ 
are the angular velocities of the grains due to radiative torques and thermal
collisions, respectively.  The radiative torques act to align grains with the 
magnetic field, while gas drag inhibits alignment and causes grains to 
point in random directions: the ratio $(\omega_\mathrm{rad} / 
\omega_\mathrm{th})^2$ therefore serves as a measurement of the effectiveness 
of the radiative torque in aligning the grains with the magnetic field.  This 
ratio will generally be highest, and the grains most aligned, in the outer 
disk where the gas density is low.  We therefore expect grains to be 
aligned in the outer disk, and oriented randomly in the inner disk. Since the 
value of $(\omega_\mathrm{rad} / \omega_\mathrm{th})^2$ varies with radial 
distance from the star, the chosen threshold value for alignment effectively 
varies the radius at which grains become aligned with the disk magnetic field.
The threshold is initially set so that grains are assumed to be aligned 
in regions of the disk where the kinetic energy imparted by the radiative 
torque is $10^3$ times greater than that imparted by thermal collisions.  
We vary this threshold in order to study its effects on the polarization 
properties of the disk. 

\begin{figure*}[ht]
\epsscale{1.1}
\plottwo{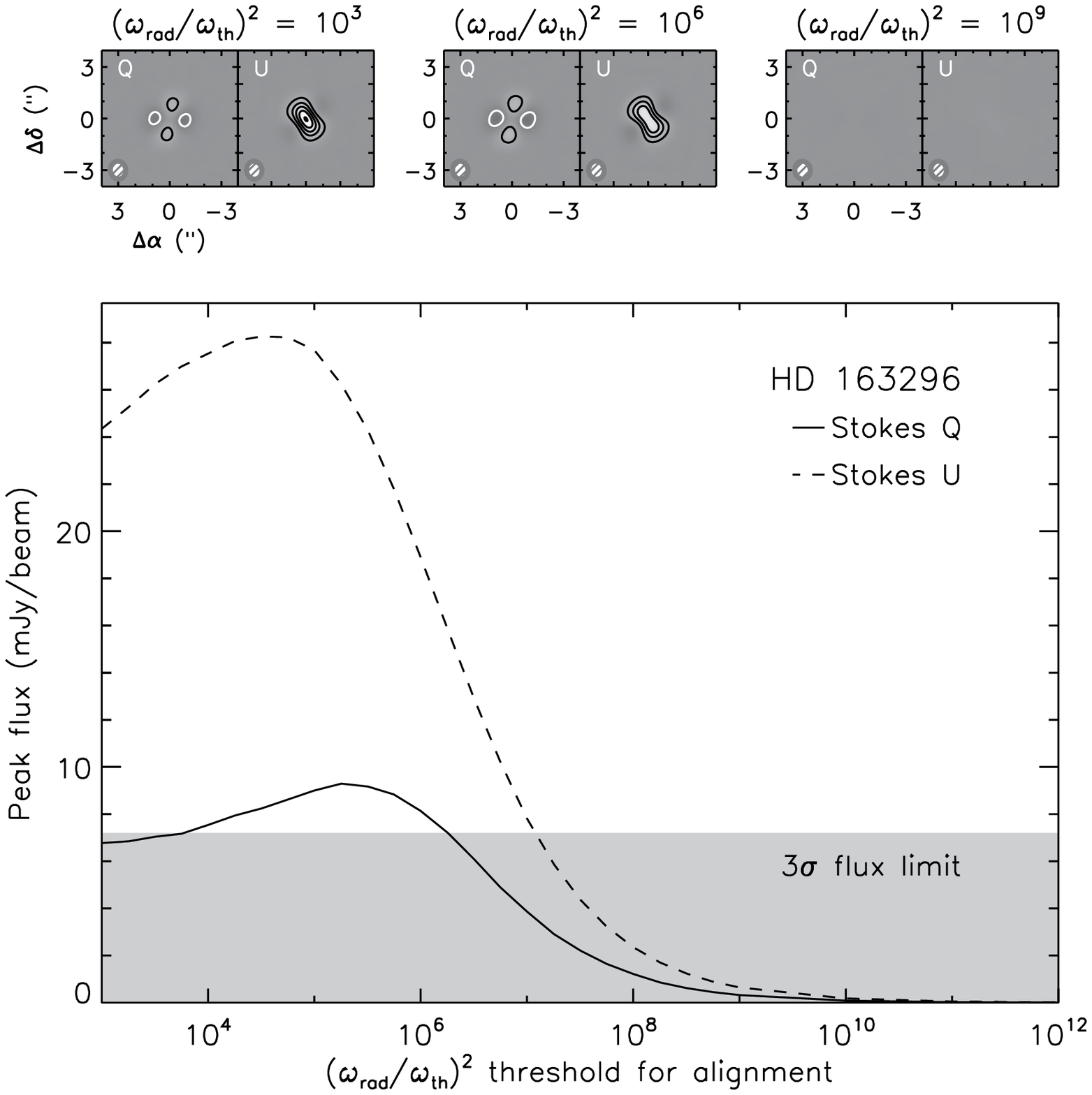}{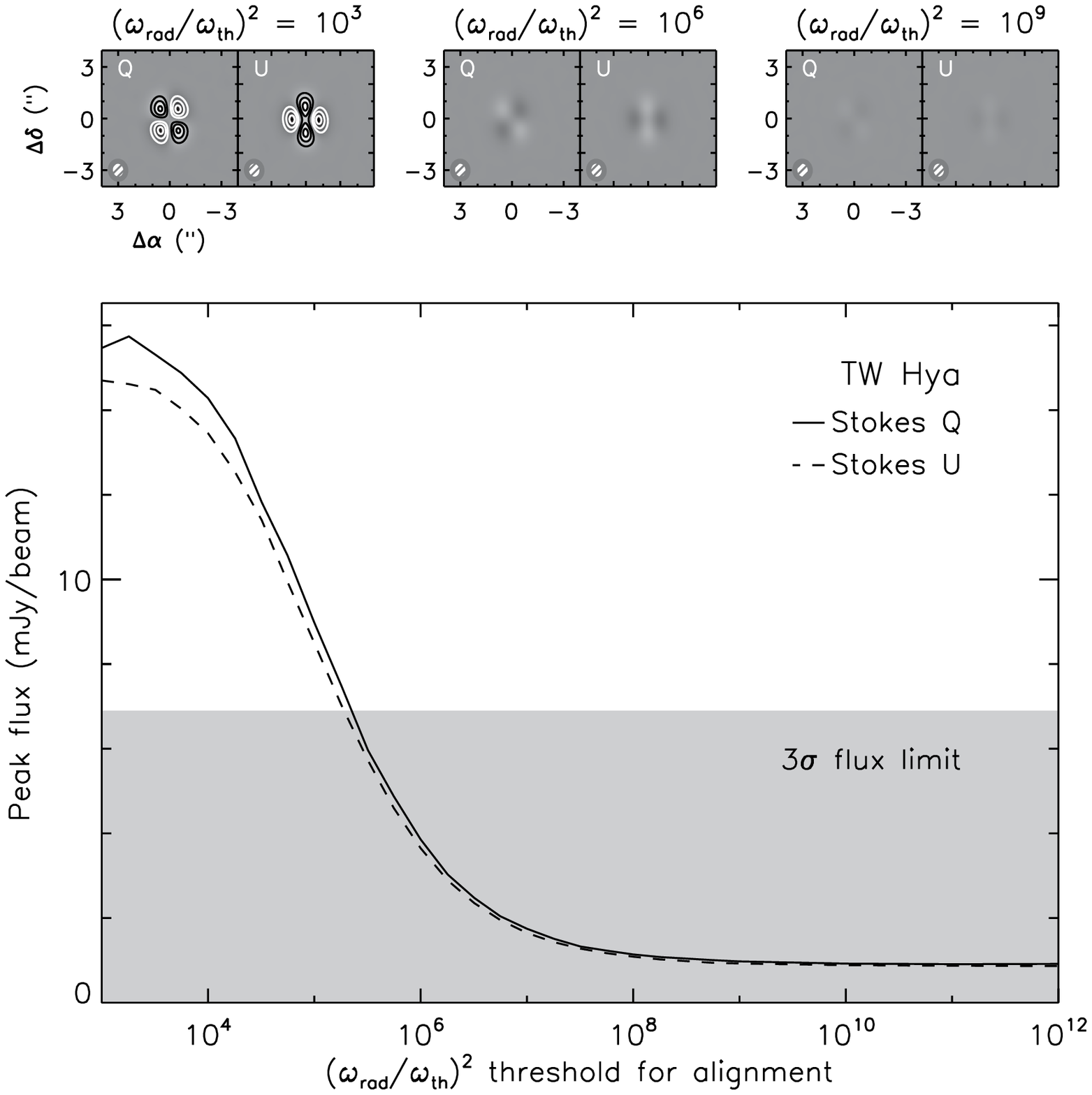}
\figcaption{
Peak continuum flux in Stokes $Q$ and $U$ as a function of the threshold for 
grain alignment (see Section~\ref{sec:params} in the text) for HD~163296 
(left) and TW~Hya (right).  The top row shows the resolved emission in Stokes 
$Q$ and $U$ predicted for three values of the alignment threshold, sampled 
at the same spatial frequencies as the data.  The grayscale indicates the 
intensity of emission relative to the peak flux of the data when the alignment 
threshold equals $10^{3}$, with white indicating positive emission and black 
indicating negative emission.  Contours are [2,4,6,...] times the rms noise 
(2.4\,mJy for HD~163296 and 2.3\,mJy for TW~Hya) with positive contours in 
black and negative contours in white.  The plots below give the peak flux in 
the synthesized beam predicted by the models as a function of the alignment 
threshold.  Stokes $Q$ is plotted as a solid line while Stokes $U$ is a 
dotted line.  The three-sigma upper limit on the peak flux from the SMA 
observations is indicated by the gray region of the plot.  The y-axis along 
the upper edge of the plot gives the dust grain axial ratio.  All images and 
peak flux values assume natural weighting to minimize noise.
\label{fig:ratioin}}
\end{figure*}

Figure~\ref{fig:ratioin} shows the peak flux predicted for Stokes $Q$ and $U$
as a function of the grain alignment threshold $(\omega_\mathrm{rad} / 
\omega_\mathrm{th})^2$, compared with the 3$\sigma$ upper limit from the
SMA observations for HD~163296 (left) and TW~Hya (right).  It is clear that
for both disks, the threshold would have to be set many orders of magnitude
higher than the conservative initial value in order for the alignment to 
be weak enough to account for the lack of a polarization signal.  Indeed, 
in order for this to be the primary mechanism suppressing the disk 
polarization, the threshold would need to be raised until alignment is
permitted to occur only when the rotational kinetic energy imparted by the 
radiative torque is at least 5-7 orders of magnitude greater than that of 
gas grain collisions.  This is most likely an unrealistically stringent 
constraint.

It should be noted here that the approach to alignment in \citet{cho07}
requires revisions to account for recent advances in the quantitative theory 
of grain alignment. First of all, in the calculations of the ratio
$(\omega_{rad}/\omega_{th})$, the simplifying assumption is made that the 
radiation seen by each grain is coming from a point source.  In fact, the
bulk of the radiation field originates as reprocessed starlight from 
neighboring regions of the disk, so although there should be an overall radial
gradient, it is better approximated by multipoles rather than a purely 
unidirectional signal.  When the effects of this radiation 
structure are accounted for, the ratio $(\omega_{rad}/\omega_{th})$ can
decrease by up to a factor of $10$ \citep[][Figure 17]{hoa09}.
An additional decrease by another factor of $\sim$10 may come from the 
fact that the overall direction of anisotropy is perpendicular to the assumed 
toroidal magnetic field in the disk \citep[][Figure 17]{hoa09}.  This effect
may be mitigated somewhat in a clumpy disk, where local anisotropies will 
not necessarily be radially oriented and may even be aligned with the magnetic
field.  Taking both effects into account and squaring the ratio demonstrates 
that the kinetic energy of the grains in their maximal state of rotation may 
be up to 4 orders of magnitude less than is assumed using ad hoc assumptions 
in the spirit of the old understanding of radiative torque alignment.  An 
additional decrease comes from the fact that an appreciable portion of grains 
may be aligned in the so-called ``zero-J'' alignment point \citep{laz07b}. 
Grains in this point are not perfectly aligned as assumed in \citet{cho07}, 
but instead will wobble, reducing the degree of alignment to only $\sim$20\% 
\citep[see][]{hoa08}. In addition, while interstellar grains are always 
aligned with long axes perpendicular to magnetic field, larger grains in 
circumstellar disks may not have efficient internal relaxation and can be 
occasionally aligned with long axes parallel to magnetic field \citep{hoa09}.  
These factors can significantly decrease the observed degree of polarization 
expected from the circumstellar disks compared to the \citet{cho07} estimate, 
making the predictions roughly comparable (to within an order of magnitude or 
so) to the SMA upper limits.

\subsubsection{Grain Size Distribution}
\label{sec:size}

\citet{cho07} emphasize the importance of the grain size distribution in 
determining the observed polarization properties of circumstellar disks.  
We fix the minimum grain size at $r_\mathrm{min} = 0.01$\,$\mu$m as in
\citet{cho07} and \citet{chi01}.  Although growth to larger sizes may have
occurred, the minimum grain size affects the millimeter-wavelength 
polarization properties in the context of the model only through the 
normalization of the total mass: increasing the minimum grain size to 
1\,$\mu$m \citep[required to reproduce the 10\,$\mu$m silicate feature from
the inner disk; 
see e.g.][]{cal02} changes the predicted polarization by less than 0.1\%, 
since it does not bring the density above the threshold value necessary
to suppress grain alignment in the outer disk.  Two aspects of 
the grain size distribution that can be varied in the context of the model 
are the maximum grain size $r_\mathrm{max}$ and the power law index 
$q_\mathrm{grain}$, where the grain size number density goes as 
$dN \propto r^{-q_\mathrm{grain}} da$.  

Observational evidence points to grain growth up to at least 1\,mm in
the HD~163296 disk \citep{ise07} and 1\,cm in the TW~Hya disk \citep{wil05},
without ruling out the possibility that grains have grown to even larger
sizes \citep[perhaps even planetary dimensions in the case of TW Hya; 
see][]{cal02,hug07}.  Since the surface density is chosen to maintain 
consistency with the observed 880\,$\mu$m flux in Stokes $I$, the number density 
of particles with sizes near 880\,$\mu$m, which dominate the 880\,$\mu$m flux,
remains roughly constant regardless of the maximum grain size in the 
distribution.  Thus the effect of raising the maximum grain size in the 
distribution is primarily to introduce ``invisible'' grains at sizes larger 
than 1\,mm or 1\,cm, which has no effect on the observable polarization 
properties \citep[cf. Figure 7 in ][]{cho07}.  However, adding mass at the 
large-grain end of the size distribution while keeping constant the mass 
in small grains has the effect of raising the total surface density of the 
disk.  This is unrealistic for all but a small increase in maximum grain 
size, as the disk quickly becomes Toomre unstable and gravitational collapse 
or deviations from Keplerian rotation should rapidly become observable.  
While this is most likely an artifact of the assumed grain size distribution, 
it suggests that within the context of the model, grain growth is unlikely 
to be the mechanism suppressing the emission of polarized radiation.

The power law index $q_\mathrm{grain}$ controlling the relative population
of large and small grains in the disk is somewhat more promising.  In general,
the polarized emission observed at a particular wavelength will tend to 
originate primarily from dust grains smaller than the wavelength, while the 
unpolarized emission will be dominated by grains of roughly the same size 
as the wavelength.  Because dust grains of size $\sim$880\,$\mu$m are within 
the geometric optics regime ($2\pi r / \lambda > 1$, where $\lambda$ is the 
wavelength of observation, 880\,$\mu$m), they do not contribute to the 
polarized emission predicted by the models.  Most of the Stokes $Q$ and $U$ 
emission at these wavelengths originates from dust grains with sizes less 
than $\lambda/2\pi \approx 100$\,$\mu$m \citep{cho07}, while most of the Stokes 
$I$ emission originates from grains with sizes similar to the wavelength of 
observation.  The relative number of 100 and 880\,$\mu$m grains in the disk, 
determined by $q_\mathrm{grain}$, therefore plays a role in determining the 
amount of polarized emission observed.  However, since the differences in 
grain sizes is not large, the power law index must change substantially 
before the effect on the polarization properties becomes appreciable.  
Varying $q_\mathrm{grain}$ from 3.5 to 2 changes the peak linearly polarized 
flux in the model by only 20\%.  Therefore, when comparing the SMA limits 
with the model predictions, the dust grain size distribution has relatively 
little impact on the predicted polarization properties of the disks.

\subsubsection{Interactions Between the Parameters}

The analysis so far has explored individual model parameters as though they 
were fully independent, determining the range of values permitted by the
SMA upper limit for each parameter separately.  However, it is useful to 
understand how the parameters relate to one another in determining the 
polarization properties of the disk.  Here we investigate relationships 
between pairs of the parameters considered above.

\begin{figure}[ht]
\epsscale{1.0}
\plotone{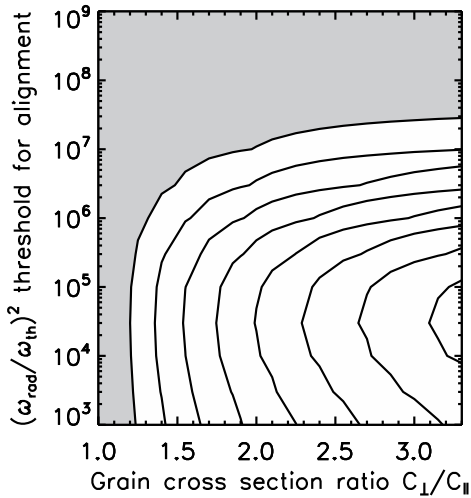}
\figcaption{
Detectability of Stokes $U$ continuum emission from the HD~163296 disk as a 
function of the dust grain cross section ratio (Section~\ref{sec:qratin}) 
and the threshold for dust grain alignment (Section~\ref{sec:ratioin}).  The 
gray regions of the plot represent portions of the parameter space that would
be undetectable given the 3$\sigma$ upper limit from the SMA observations, 
while contours show the peak flux of the model for each set of parameters, 
beginning at the 3$\sigma$ level (7.2\,mJy) and increasing by intervals of 
2$\sigma$ (4.8\,mJy).  The two parameters are only weakly degenerate.
\label{fig:twod}}
\end{figure}

We first study the relationship between dust grain elongation and 
the grain alignment threshold.   As discussed in \S2.3 of \citet{cho07}, the 
rotation rate of dust grains due to the radiative torque is a function of 
the peak wavelength of the radiation field and the dust grain size, with no 
explicit dependence on grain axial ratio.  As described in \citet{dol76}, 
spin-up by the radiative torque mechanism is caused by the irregular shape 
of the grain, which gives it differing cross sections to left and right 
circular polarization; elongation does not  necessarily favor either 
polarization basis.  This is reflected in the table of timescales relevant 
for grain alignment in \citet{laz07}: neither the radiative precession time 
nor the gas damping time depends on the grain  axial ratio.  The primary 
effect of the grain elongation in alignment is to decrease the Larmor 
precession time, which causes the spinning grains to align their major axes 
more quickly with the magnetic field lines (or, alternatively, decreases 
the critical magnetic field strength in a given region of the disk; see 
Section~\ref{sec:other} below).  We therefore do not expect much, if any, 
dependence between these variables.  In order to test this expectation, 
we vary the dust grain cross section ratio and the grain alignment threshold 
for the HD~163296 disk.  The model prediction of peak flux in Stokes $U$ 
(which provides the most stringent limits when compared to the SMA data) 
are shown in Figure~\ref{fig:twod}.  The shaded gray region of the 
plot represents the parameter space within which the model prediction is 
less than the 3$\sigma$ upper limit given by the SMA data, i.e., combinations 
of parameters consistent with the observational results.  The contours show 
the predicted peak flux of the model in Stokes $U$ for each combination of 
parameters: model predictions with greater polarized intensity are more 
strongly inconsistent with the observational limits.  Because of the 
assumption in the models that grains meeting the alignment criterion will 
become aligned with 100\% efficiency, grain alignment and elongation are 
evidently only weakly coupled.

Another potentially important relationship is that between the grain
size distribution and grain elongation.  Little is known about the 
relationship between these variables, since both are notoriously 
difficult to constrain observationally.  Nevertheless, if grains grow simply
by accumulating material evenly over their surface then they may naturally 
become more spherical as they become larger.  Spherical grains emit less 
strongly polarized radiation than more elongated grains, so it might be 
expected that grain growth can suppress the emission of polarized light, 
even in cases where the alignment mechanisms are quite efficient \citep[as
expected for large grains, e.g.][]{cho05}.  Indeed, a corresponding inverse 
relationship between grain size and polarization fraction has been observed 
in molecular clouds \citep[e.g.][]{vrb93}.  Given the observed growth to 
millimeter and even centimeter sizes within the disks around HD~163296 and 
TW~Hya \citep{ise07,wil05}, and the large ($\sim$100\,$\mu$m) sizes of the 
grains responsible for emitting most of the polarized radiation (see 
Section \ref{sec:size}), it is perhaps plausible that the grains in 
these disks should have cross section ratios consistent with the values of
1.2-1.3 constrained in Section~\ref{sec:qratin} above.  We know that this 
cannot be true everywhere in the interstellar medium (ISM): polarization at 
850$\mu$m is observed in star-forming regions at much earlier 
stages \citep[e.g.][]{gir06}, and far-infrared polarimetry indicates that 
grains with axial ratios $a/b$ between 1.1-3 are common at sizes of tens of 
microns in molecular clouds \citep{hil95}.  However, a tendency towards
spherical grains in T Tauri disks, even just at the low end of the 
distribution inferred by \citet{hil95}, should be able to suppress the 
emission of polarized radiation from the disk enough to bring the models 
within range of the observational constraints.

We can test the plausibility of this degree of elongation by modifying the
discussion of grain growth based on turbulent coagulation in \citet{vrb93}.  
If we assume that the grains in T Tauri disks originate exclusively from 
small, highly elongated grains in the ISM, e.g. with initial major axis 
$a_i = 0.1$\,$\mu$m and axial ratio $a_i/b_i = 2$ \citep{aan73,hil95}, then 
we can estimate how the axial ratio changes with grain 
size.  Neglecting asymmetric effects like collisional destruction, 
grain size might be expected to grow roughly evenly in all directions 
with the number of grain-grain collisions, $N$, in such a way that the final 
grain size is simply $a_f = a_i N^{1/3}$.  The change in any dimension of 
the grain, $\delta$, is then given by $\delta = a_i N^{1/3} - a_i$, yielding 
a final minor axis size of $b_f = b_i + \delta$, or $a_f/b_f = a_f/(a_f - 
a_i+b_i)$.  If $a_i$ = 0.1\,$\mu$m and $a_f$ = 100\,$\mu$m, then $a_f/b_f$ = 
1.001, significantly more round than the upper limit set by the SMA data.  
The timescale needed for grain growth to these (up to meter) sizes is of order 
$10^5$ years at a distance of 50\,AU from the central star \citep[see 
e.g.][]{wei88,dul05}.  This calculation is highly simplified and neglects 
complications like the evolution of conditions within the disk, shaping by 
grain-grain collisions \citep[e.g.][]{dul05}, and the complexity of the 
grain size distribution.  Yet the extremely spherical grains produced on 
relatively short timescales in this oversimplified scenario represent a 
lower limit to the grain elongation that perhaps suggests a scenario by 
which grains might have grown into shapes that are nearly spherical enough 
(with axial ratios of 1.2 rather than 1.001) to plausibly account for the 
suppression of polarized emission.

\subsection{Other Effects}
\label{sec:other}

In the previous section, we investigated the effects of those parameters 
considered in the \citet{cho07} model.  However, there are additional 
effects that may also play a role in suppressing polarized emission from 
the disk relative to the fiducial 2-3\% prediction.  Among these are the 
magnetic field strength, the geometric regularity of the magnetic field, 
and polarization due to scattering.

\subsubsection{Magnetic Field Strength}

The magnetic field strength plays a role in determining whether or not 
grains can become aligned via the radiative torque.  If the magnetic field is
above some critical strength, grains will become aligned provided that 
the radiative torque can generate more rotational kinetic energy than thermal
collisions.  At low magnetic field strengths, grains are not expected to 
align with the magnetic field at all.  The critical magnetic field strength 
for alignment may be estimated by comparing the Larmor precession time $t_L$ 
with the gas damping time $t_{gas}$.  Following \citet{laz07} and using 
fiducial values for the magnetic susceptibility and dust grain density, 
alignment is possible when $t_L<t_{gas}$, or:
\begin{eqnarray}
B > 4.1 \times 10^{-5} \frac{r n T_d T_g^{1/2}}{s^2} 
\end{eqnarray}
where 
$B$ is the magnetic field strength in units of $\mu$G, 
$r$ is the grain size in cm, 
$n$ is the gas density in units of cm$^{-3}$, 
$T_d$ is the dust temperature in K, 
$T_g$ is the gas temperature in units of K, and  
$s$ is the ratio of minor to major dust grain axes.
Using the power law models of 
density and temperature derived in \citet{hug08}, it is possible to estimate 
these quantities for the regions of the outer disk probed by the SMA data.  
Taking the values at disk radii equivalent to the spatial resolution of the 
data ($\sim$1\farcs0, or 50 and 120\,AU for TW~Hya and HD~163296, 
respectively), and assuming equivalent gas and dust temperatures, we derive 
densities of several times $10^8$\,cm$^{-3}$ and temperatures of $\sim 
40-50$\,K.  For the 10-100\,$\mu$m grains contributing most of the polarized 
emission in the models, the critical magnetic field strength is of order 
10-100\,mG. 

This strength matches reasonably well with theoretical expectations.  
\citet{shu07} developed a model of steady-state magnetized accretion disks
that predict magnetic field strengths of order 10-100\,mG on the spatial
scales probed by the data.  \citet{war07} pointed out that Zeeman splitting
of OH in molecular cloud cores and masers in star forming regions place a 
lower limit of $\sim$10\,mG on the magnetic field strength, which will 
likely be amplified by compression and shear during the process of collapse
that forms the central star and disk.  It should also be noted that the 
value quoted above should be taken as a lower limit, since superparamagnetic 
inclusions would significantly decrease the required magnetic field strength 
for alignment \citep{laz08}.  The critical magnetic field strength required 
to align grains within the conditions of the model is therefore reasonable 
compared to theoretical expectations.  We do not expect that the lack of 
polarized emission is due to extremely low magnetic field strengths.

\subsubsection{Geometric Regularity of the Magnetic Field}

The assumption that the field is toroidal arises from the supposition
that the rotational motion of the disk has affected the magnetic field
geometry.  Yet for this to occur, the ionization fraction must be large
enough that disk material and magnetic fields can interact.  However, this
also implies that turbulent motions within the disk (perhaps even of magnetic
origin) may tangle the magnetic fields locally, adding a random component
to the ordered toroidal magnetic field.  It is extremely difficult to 
estimate the magnitude of such an effect without knowing both the ionization
fraction and the magnitude of turbulence as a function of position in the
disk.  \citet{lee85} discuss the effect of a random magnetic field component
on the strength of the observed polarization signature, and note that the
strength of polarized emission will be reduced by a factor $F = 3/2 ( 
\left\langle \cos^2 \theta \right\rangle - 1/3)$, where $\theta$ is the angle 
between the local magnetic field and the direction of the ordered global 
magnetic field.  This quantity varies from one (perfectly ordered field; 
$\left\langle \cos^2 \theta \right\rangle = 1$) to zero (perfectly random 
field; $\left\langle \cos^2 \theta \right\rangle = 1/3$), but the exact value 
depends on the details of the local magnetic field geometry.  If magnetic 
field tangling were the sole factor responsible for the difference between 
the fiducial modeling prediction and the SMA upper limits, we would constrain 
$F$ to be less than $\sim0.1$ for the case of HD~163296, implying 
$\left\langle \cos^2 \theta \right\rangle < 0.4 $, which indicates an almost 
completely random magnetic field structure.

It should also be noted that grain alignment efficiency would play a similar
role, quantified in exactly the same way as $F$ above, with $\theta$ 
indicating the angle between the long axis of the grain rather than the 
angle between the local and global magnetic fields \citep{gre68,lee85}.  
The \citet{cho07} code assumes 100\% efficient alignment in regions that
meet the grain alignment criterion (Section~\ref{sec:ratioin}).  In order
to account fully for the suppression of polarized emission relative to the
fiducial model, the alignment efficiency would have to be quite low, less
than 10\% in the case of HD~163296.

The tentative single-dish detections appear to indicate a toroidal
magnetic field geometry for the disks around DG Tau and GM Aur, consistent
with observations indicating a dominant toroidal component to the magnetic
field in the flattened structures around YSOs at earlier evolutionary stages 
\citep[see][and references therein]{wri07}.  However, it is also possible 
that the field could be poloidal: as discussed e.g. in \citet{shu07}, a 
magnetic field gathered from the interstellar medium that threads vertically 
through the disk might be expected to remain poloidal in geometry as it 
interacts with disk material.  While the SMA limits are unable to constrain 
the magnetic field geometry, a poloidal geometry might be expected to reduce 
the expected polarization signature particularly for the case of a face-on 
viewing geometry as in the case of the TW Hya disk.  The effects of a poloidal 
geometry for a disk viewed at intermediate inclination, like HD 163296, 
are less clear and are not investigated in the context of the \citet{cho07} 
models, although it is plausible that the strength of polarized emission 
from a toroidal or poloidal field would be comparable.

\subsubsection{Scattering}

\citet{cho07} argue that scattering contributes significantly less than
thermal emission to the polarized flux at millimeter wavelengths in the disk.  
In order to estimate the relative contribution of scattering and emission at
a range of radii throughout the disk, they compare the product $J_\lambda 
\kappa_\mathrm{scatt}$, where $J_\lambda$ is the mean radiation field and 
$\kappa_\mathrm{scatt}$ is the mass scattering coefficient, to the product 
$B_\lambda \kappa_\mathrm{abs}$, where $B_\lambda$ is the intensity of 
blackbody radiation in the region of interest and $\kappa_\mathrm{abs}$ is 
the mass absorption coefficient.  They show that in the outer disk, where 
$R \gtrsim 10$\,AU, the ratio of these products falls below one (and ultimately
below 0.5), indicating that emission is dominant over scattering in the 
outer disk.

It is of interest, however, that pure scattering of light from a central 
source off of large grains in the outer disk should produce a polarization 
signal precisely orthogonal to that expected for elongated grains aligned 
with a toroidal magnetic field.  While the radiation field at 850\,$\mu$m 
is dominated by the local conditions rather than a central source, as 
discussed in Section \ref{sec:ratioin} there will be an overall radially 
anisotropic component of the radiation field that might be expected to 
produce a weaker, but still orthogonal on average, scattering signal.  The 
contribution from scattering would therefore generally act to cancel the 
expected polarization signal from emission.  An estimate of the magnitude 
of the scattered light signal compared with the predicted strength of polarized 
emission is beyond the scope of this paper, but we note that for scattering 
to be the dominant mechanism suppressing the expected polarization signal, 
the intensity of polarized emission arising from scattering and emission 
would have to be precisely equivalent, to within 10-15\%, in both disks.  
Furthermore, since the scattering and emission have different wavelength 
dependences, the coincidental canceling of the emission signal would only 
occur at the wavelength of observation.  In the absence of any expectation 
that these quantities should be related, this seems an unlikely coincidence.

\section{Summary and Conclusions}
\label{sec:sum}

Despite the expectation of a 2-3\% polarization fraction in circumstellar
disks based on previous observational and theoretical work \citep{tam99,cho07},
the SMA polarimeter observations presented here show no polarization from
the disks around two nearby stars.  With these observations we place a 
3$\sigma$ upper limit on the integrated polarization fraction of less than 
1\% and rule out the fiducial \citet{cho07} models at the $\sim$10$\sigma$ 
level.  These represent the most stringent limits to date on the magnitude 
of submillimeter polarized emission from circumstellar disks.  We are 
therefore left with the question of which model assumptions are unrealistic 
enough to account for an approximately order-of-magnitude (at minimum) 
overprediction of the polarization signal from these disks.  

Among the model parameters and additional effects considered in 
Section~\ref{sec:analysis}, several seem unlikely as the source of the
suppression of polarized emission.  The critical magnetic field strength 
expected for alignment seems reasonable relative to theoretical expectations
and observations.  An almost completely random magnetic field with 
no dominant toroidal (or poloidal) component would also be surprising, although
a poloidal field geometry would be expected to significantly weaken the
polarized emission arising from a face-on disk like TW Hya. 
Scattering is expected to be weak, but it should produce a polarization 
signature perpendicular to that expected for emission from aligned grains. 
However, scattering and emission signals would have to cancel nearly perfectly 
in order to account entirely for the low observed polarization fraction.  
Nevertheless, there are promising candidates to describe how the suppression 
of polarized emission might have occurred.  \citet{cho07} assume 100\% 
efficient alignment of grains with the magnetic field in regions of the disk 
where the alignment criterion is met, which is overly optimistic and now known 
to be unrealistic (see discussion in Section~\ref{sec:ratioin}).  In light of 
the recent work on the quantitative theory of grain alignment \citep{laz07b,
hoa08,hoa09}, the \citet{cho07} result may be considered an upper limit to the 
theoretical expectation for the polarization properties of disks.  A reduction 
to 10\% efficiency, which is within the expectations based on recent 
developments in grain alignment theory, could alone explain the low 
polarization fraction observed.  Another possibility is that the grains 
contributing most of the polarized emission in the model are well (or not so 
well) aligned, but rounder than the cross section ratio assumed in the initial 
model and therefore inefficient emitters of polarized radiation.  This is
also reasonable based on a rough estimate of the timescales and shapes 
expected for collisional growth of elongated ISM grains.  

While each of these factors would have to be substantially different from
what is expected in the initial model to alone account for the low polarization
fraction, it is of course entirely possible that several effects are playing
a combined role.  For example, grains with a cross section ratio of 1.5 
instead of 2.1 could combine with a 50\% alignment efficiency to account 
entirely for the difference between observations and models.  A small degree 
of field tangling (expected because of turbulence in the disk) could further 
reduce the expected polarization signature.  While we cannot constrain 
precisely which factors are contributing in which proportions to the 
suppression of polarization in the disks observed with the SMA, we identify 
these three factors (grain elongation, alignment efficiency, and field 
tangling) as the most plausible sources of the suppression of polarized 
emission.  They produce the greatest change in polarization properties 
within a reasonable range of parameter values, and there exists a theoretical 
justification for why they should exist, even if the magnitude of the effect
is not well constrained.

Future observations with higher sensitivity may be able to disentangle these
effects to some extent, particularly the degree of field tangling.  It would
also be useful to obtain high spatial resolution observations of the
disks with tentative detections of a 2-3\% polarization fractions to 
confirm the strength and origin of the emission on small spatial scales, 
and to expand the sample size in order to determine whether the low
polarization fraction constrained by the SMA is universal for disks around 
young stars.  

\acknowledgments{We thank Alyssa Goodman for helpful conversations that 
improved the manuscript.  Partial support for this work was provided by 
NASA Origins of Solar Systems Program Grant NAG5-11777.  A.~M.~H. 
acknowledges support from a National Science Foundation Graduate Research 
Fellowship.  Support for S.~M.~A. was provided by NASA through Hubble 
Fellowship grant \#HF-01203-A awarded by the Space Telescope Science 
Institute, which is operated by the Association of Universities for Research 
in Astronomy, Inc., for NASA, under contract NAS 5-26555.}

\bibliography{ms}

\end{document}